\documentclass[acmsmall,preprint,authorversion]{acmart}

\setcopyright{none}
\usepackage{booktabs}   
\usepackage{subcaption} 

\newcommand{\RED}[1]{\textcolor{red}{#1}}
\newcommand{\eg}{\emph{e.g.,}}

\newcommand{\anonymize}[2]{#1}

\makeatletter\if@ACM@journal\makeatother
\acmJournal{PACMPL}
\acmVolume{1}
\acmNumber{1}
\acmArticle{1}
\acmYear{2017}
\acmMonth{1}
\acmDOI{10.1145/nnnnnnn.nnnnnnn}
\startPage{1}
\else\makeatother
\acmConference[PL'17]{ACM SIGPLAN Conference on Programming Languages}{January 01--03, 2017}{New York, NY, USA}
\acmYear{2017}
\acmISBN{978-x-xxxx-xxxx-x/YY/MM}
\acmDOI{10.1145/nnnnnnn.nnnnnnn}
\startPage{1}
\fi

\setcopyright{none}             

\usepackage{cite}
\usepackage{placeins}

\usepackage{cite}
\usepackage{breakurl}
\usepackage{xspace}
\usepackage{dsfont}
\usepackage{amsmath,amsthm,amssymb,amsfonts,mathtools}
\usepackage{listings}
\usepackage{wrapfig}
\usepackage{graphicx}
\usepackage{enumerate}
\usepackage{framed}
\usepackage{multirow}
\usepackage{adjustbox}
\usepackage{paralist}
\usepackage{xcolor}

\usepackage{rotating}
\usepackage{pifont}

\definecolor{light-gray}{gray}{0.85}
\definecolor{Gray}{gray}{0.9}
\definecolor{darkblue}{rgb}{0,0,.75}

\title{Mastery Learning-Like Teaching with Achievements}

\author{Tobias Wrigstad}
\affiliation{%
  \institution{Uppsala University}
  \country{Sweden}
}
\email{tobias.wrigstad@it.uu.se}

\author{Elias Castegren}
\affiliation{%
  \institution{Uppsala University}
  \country{Sweden}
}
\email{elias.castegren@it.uu.se}

\begin{document}

\begin{abstract}
  This paper describes the design of a second-year, 20 ECTS credit
  course on imperative and object-oriented programming. The key
  design rhetoric is encouraging students to assume responsibility
  for their own learning, and, to this end, elements from mastery
  learning and other teaching strategies are used in concert to
  construct an achievement-based system where students are put in
  charge of how and when they are examined. In addition to
  describing the elements of the course design, the paper reports
  on experiences from teaching in this format, and how the format
  has evolved over time.
\end{abstract}



\maketitle

\section{Introduction}
\label{sec:introduction}

Four years ago at \anonymize{Uppsala University}{University redacted}, we changed the 2nd year 20
ECTS credit course\footnote{An ECTS credit is a standard EU credit unit. 20 ECTS credits corresponds to $\approx$533 hours of work.} that introduces imperative and object-oriented
programming, with focus on coding and tools. With the new course design, we hoped to change the
way students approach learning and also force students to assume
more responsibility for their studies. This paper summarises the
outcomes of this undertaking: it explains the key course design
elements, how the course design has changed from its initial
implementation to its current format, lessons learned from the
process, and a biased, qualitative self-evaluation.

First and foremost, it is important to understand the setting in
which this takes place: Sweden, a country where university education is
``free'' in the sense that there are no tuition fees and students
are given a basic allowance and state-sponsored loans for
studying. The course in question, \emph{Programming Methodology
  for Imperative and Object-Oriented Programming}, has 120--140
students yearly and is a mandatory 2nd-year course on both the
traditional computer science bachelor programme and the five-year
engineering programme in information technology. Before this course,
students have been taught functional programming (lately with
Haskell), basic data structures, program construction principles,
and basic computer architecture including a little assembly language.
Each batch of students can be divided into 3 roughly equi-sized groups ---
those who are already programmers (programming extracurricularly),
those who learned imperative programming in upper-secondary
school, and those who have only studied programming at the
university. The course runs for a full semester at 66\% speed ($\approx$ 530 working hours for each student); 
students take one other course in parallel first half of the
semester, and another in the second half. These courses vary
with degree program but are usually math-oriented.

In our experience of programming courses (not just limited to
\anonymize{Uppsala}{Our university} or \anonymize{Sweden}{Our country}), student assessment usually comes in the shape
of a number of implementation-based assignments. These assignments are
often cleverly designed so that concepts just taught in lectures
can be successfully employed to solve the problems at hand.
A student is thought to have mastered the key concepts if she is
able to provide a ``working implementations'' of the assignments
that pass some key tests.

We find this practice problematic for several reasons:

\begin{enumerate}[a)]
\item it rewards students' ability to pattern match recent lecture
  content onto a problem which has little bearing on
  ``real-world'' programming, and rewards the ability to solve
  programs the way the teacher intended rather than using independent
  thinking;

\item as students commonly collaborate, there is no real check
  that a student has worked on all parts of a program and thus
  actively reflected on all goals of the assignment;

\item because the problem of implementing a program that performs
  $X$ is so much more concrete than grasping abstract topics like
  ``encapsulation'', the former tends to overshadow the latter. Many
  predominantly qualitative aspects of software, like \eg{}
  maintainability, are hard to understand without living with code
  for a very long time and suffering from own mistakes/choices.

\end{enumerate}

We set out to develop a course that helps students reflect on
skills and actively think about what they are doing, and where we
could be reasonably sure that all students were actively engaging
not only with the practice of programming, but also with more
general concepts that go beyond the ``random'' set of concepts
that surface in the parts of a program a particular student
happens to write. At the same time, we wanted to cater equally to
the above-mentioned groups of students, allow experienced programmers to
quickly dive into more complicated matters, and leverage the fact
that they are mostly self-serving in a way that would help us
direct more resources to those with less programming experience.
Also, in particular, we wanted to make the students actively make
decisions about their learning, as well as what to learn when,
rather than just sit down and ``be taught''. In the process, we
developed a system we came to call ``Achievement Unlocked'', which is
similar to mastery learning
\citep{bloom1968learning, guskey2010lessons}, although we had not
heard about that at the time. Our main source of inspiration was
constructive alignment \citep{biggs1996enhancing}. In this
paper we report on the use of Achievement Unlocked in this one
particular course. In future papers, we hope to explore the key
ideas more deeply, connect them to existing theories like mastery
learning, and discuss their application to teaching advanced
software design.

\paragraph{Outline}

Section 2 describes the course structure, the achievement system,
and how students demonstrate mastery, plus the interplay with \eg{}
lectures, assignments, our version of a closed-book exam, and how
to give a final grade to each student. Section 3 reports on our
self-evaluation, based to a large extent on feedback from students
and teaching assistants. Finally, Section 4 concludes.

\section{Course Structure}
\label{sec:course-structure}

As this course fits in two larger degree programmes, it must serve many
masters, and has thus been subject to ``feature creep'' in its
formal course goals over the years. In addition to learning how to program
in imperative and object-oriented languages, students are expected
to learn about programming methodologies and tools, as well as basic
software engineering, in particular agile software development
processes. Additionally, soft skills such as oral presentation and
report writing should be taught, as well as professional ethics (to a limited extent).
An overview of the achievements, a sample achievement description and the course outline with syllabus, timings and deadlines can be found in Appendix.

\subsection{The Achievement System}
\label{sec:achievements}

The course is centered around a relatively large set of
\emph{achievements}, around 70, which are derived from the formal course
goals and checked against the ACM/IEEE Curriculum for Computer
Science \citeyear{acmcurriculum}. Each achievement belongs to a certain grade level---3, 4
or 5---and to pass with a certain grade, a student must pass 100\%
of the achievements belonging to that grade, and those below. Some
achievements only exist on the basic grade level (3) whereas
others exist at increasing depth across all grade level. An
example of the former is ``writing documentation''\footnote{Not
  because there is not enough depth in the subject, but because of
  prioritising other subjects.}. An example of the latter is
profiling, where running a profiler, producing profiling output and
using it to explain a program's performance is at grade level 3,
using profiling output to improve and demonstrate the performance
improvement is at level 4, and level 5 includes reducing memory
footprint.

Achievements are skills that students must demonstrate ``mastery
of'' at least once during the course in order to pass the course.
Importantly, achievements are not tied to any specific assignments
or dates. Instead, students are required to \emph{map the
  achievements} to activities during the course (for example,
programs they write as part of mandatory assignments) and may
\emph{attempt to demonstrate} mastery when they so choose. Section
\ref{sec:demonstrating} discusses demonstration in
further detail.

Because any achievement can be demonstrated at any time, students
are free to attempt to demonstrate mastery when they feel they
have ``grokked'' something, not at an externally-controlled moment
in time. Some students demonstrate piecemeal interleaved with
assignment work, others first complete an assignment and then make
fewer but more comprehensive demonstrations. Students
are strongly encouraged to make demonstrations that tell
a coherent story. A student that realises the
connection between ``identity and equality'' and ``aliasing'' will
be able to demonstrate both of these achievements together, and in
less time than it would take to demonstrate them both separately.

\subsection{Influencing Student Behaviour Through Resource Limitation}
\label{sec:contr-stud-behav}

While there are abundant opportunities to demonstrate, the number
of lab sessions with demonstration slots are limited, as are the number of achievements
that can be demonstrated in an attempt and even successfully demonstrated in a session. This is an integral part
of the course design. Making the demonstration attempts a limited
resource increases their value, and discourages ``waste'' in the
sense of attempting to pass by brute force (because students
invariably learn something from failed demonstration attempts).
The last two times the course was given, there have been
around 30 slots for presenting, and a maximal number of 4
achievements per demonstration.\footnote{About 60 of the 70 achievements
  are demonstrated in lab sessions. Thus, a student could theoretically successfully demonstrate enough achievements to pass the course twice with the highest grade, or fail 50\% of all attempts and still pass with highest grade, within the allowed number of demonstration attempts.} This encourages
\emph{clustering of achievements} to form larger presentations.
Students unable to form synthesis are at a disadvantage as they
will not be able to cluster achievements as effectively. This is
on purpose as the ability to form synthesis is something that we
wish to have reflected in a student's grade. Students say that the
emphasis on clustering forces them to work with the subject
continuously throughout the course (from course evaluation and interviews).

To discourage students abandoning good practices after
successfully demonstrating mastery, we randomly select
achievements from a student's collection of passed achievements
for re-examination. The re-check is intended to be superficial,
and mostly serve as a means of jogging students' memory and
keeping knowledge fresh. A successful demonstration cannot be
``lost'' if a re-check fails, but the student will be re-checked
again until the re-check succeeds. This way, re-checks can pile
up, effectively blocking students from making progress with other
demonstrations.

\subsection{Achivement Choice \& Specification}
\label{sec:achivement}

The actual selection of achievements seem not to be crucial to the
course design and we have tweaked both the number, distribution
and content of the achievements over the last years. Our course
focuses on imperative and object-oriented programming, and
emphasises methodology and tools, which is reflected in the
achievements. For example, we have achievements related to using
version control, using the Unix shell, build management and source code
editing.

As part of the goal of forcing students to work with achievements
is to force them to make their own plans and map achievements to
activities in the course, we have experimented with presenting the
achievements in list form, where achievements belonging together
were grouped, but without highlighting this or giving names to the
groups. This seems to overwhelm some students who panic at the
sheer volume, and who quickly filter out achievements connected to
higher grades simply to make the list more manageable. Our current
design is to give each group a name and present the achievements
in this group together. It seems that the grouping of assignments
in named groups is more important than the actual groups, likely
because it makes the set of achievements less daunting for the students.

As the course material is currently only available in \anonymize{Swedish}{a language that is not English}, we
provide a list of the groups of achievements used in 2016, and a
rough outline of the achievements in each group in the Appendix,
together with an example achievement.

To simplify the course structure, we try to fit everything inside
the achievement system, even though that sometimes goes against
the  idea of achievements. Thus, we represent each completed assignment as an
achievement, as well as the final project. This does not seem to be
confusing to students, but possibly comes at a cost of making the
achievements idea less pure. At the end of
the course, there are often students who have demonstrated all
achievements except one or more complete assignments.
To us, this highlights the importance of decoupling
skill demonstration from \emph{finished} artefact implementation.
By giving specific assignments, we ensure that students write substantial
amounts of code in the course,\footnote{Throughout the course,
    the average student writes a
    total of 4--5.000 lines of code, not including introduction labs,
    extensions of assignments to address a certain achievement, or
    the course project.}
and that some parts of the code survive
long enough that the pain of bad choices, bad naming, etc. can be learned from.

\subsection{Demonstrating Mastery}
\label{sec:demonstrating}

With few exceptions, all demonstrations take the form of oral
examination in front of a computer terminal with one examiner and
two students working in a pair. There are several reasons for
pairing students together: the students outnumber the examiners,
reducing stress and taking pressure off the situation; if a
student gets stuck, the partner can take over or help out; it is
easier to create a meaningful discussion with more than two
people---for example, the examiner can ask the other student
whether she agrees or disagrees with a statement that the first
student made, or if she can propose a different way of doing
something than what the first student just explained.
Examining two students at a time also cuts the total number of
demonstrations in half, saving examiner time.

Checks are individual, meaning that both students must
demonstrate mastery. A mostly passive student will
not pass a check even with a brilliant partner. It is clearly
stated that the responsibility for making sure that both students
actively participate is on the students, and not on the examiner.
This encourages students to prepare the demonstration and
discuss who says what, which in our experience makes
presentations run faster, have higher quality, and increase
student reflection.

The choice of oral examination serves several purposes. Explaining
interactively is time-efficient, both for the student and
examiner, and feedback is given immediately when the demonstration
is still fresh in the student's mind. Technical communication is an
important soft-skill that is trained automatically in this
setting, and students are both encouraged and helped to improve
their ability to communicate about software development. Nevertheless, some
students complain, saying that they are not interaction-oriented
and much prefer to explain themselves in writing. We explain to
  them that communication and interaction are key skills, and do not
  provide any alternative means for them to demonstrate.

\subsubsection{Demonstration Process}
\label{sec:demonstr-proc}

Demonstrations take place in designated computer labs and (for
students with laptops) lecture halls, during 4 hour lab sessions, which
are scheduled twice weekly throughout the semester.
Students request to demonstrate through a web page where they
choose which achievements they wish to demonstrate. An achievement
for re-check is added automatically by the system, using the
history of passed demonstrations. Re-checks are disabled early in
the course, until students have passed a sufficiently large
number of assignments. On average 7--8 TAs serve 100 students in each
session. (About 3 senior TAs and 10--12 junior TAs are attached to the course.)

After requesting to demonstrate, students wait (ideally while continuing to work). 
A feed of pending requests to demonstrate is shown on a screen to the examiners,
who claim demonstrations before dispatching to where the students
are located. We follow no special rules for claiming
demonstrations: sometimes it makes sense to have the same examiner
claim a student's second attempt, other times it makes sense to
hand over to someone else. To increase throughput, it sometimes
makes sense to have one examiner claim several demonstrations in
the same room. As the achievements a students wishes to demonstrate
is visible in the feed, examiners can avoid achievements they feel
uncertain about, request help from a more senior examiner, etc. 

Students are notified that their request has been picked up and by whom. At
all times, students can see how many requests are waiting and how
many are in front of their request, if any. Students are not
allowed to have more than one pending request or make requests
before lab sessions start.

When the examiner arrives, the students must \emph{pitch their
  demonstration}, which means:
\begin{enumerate}
\item Stating the achievements they wish to demonstrate
\item Explaining why/how it makes sense to demonstrate them together
\item Explaining what evidence they will use in the demonstration
\end{enumerate}

If the examiner is not satisfied with 1 or 2, she
will explain why and should reject the request (in
practice, this has not been strictly enforced as we want to avoid
discouraging students from \emph{trying} to pitch a demonstration).
By evidence in 3, we mean artefacts developed in the course that
will form the basis of the demonstration. For example, ``we
profiled the binary search tree that we developed in Assignment
2''. The key idea with evidence is that students must relate the
achievements to the programming assignments (and programming
project) of the course. In the course's initial instalment, we
naively allowed any code to be used in demonstrations prompting
students to search for code rather than write their own, or
spending lots of time trying to write a minimal program that could
be used to demonstrate a maximal number of achievements. While
there are merits to both these approaches, we are seeking to teach how to write
imperative and object-oriented code, which is why we added the
requirement of evidence in the form of assignments from the course.
Examiners will often ask questions to force students to go
``off-script'', \eg{} ask what would happen if $X$ was changed for $Y$,
or change some lines of code, or introduce a bug etc.

If the examiner finds the answer to 1--3 satisfactory, she will
hand over the running of the demonstration to the students.
Different examiners approach this task differently. Some will
interrupt and ask questions. Some delay all questions to the
end. What approach students prefer seems highly subjective, with a
slight preference for more interactive examiners.

Only the achievements explicitly stated at the beginning of the
demonstration can be passed in the
demonstration. If a student in the process of demonstrating
happens to ``meet the requirements'' for some other achievement, this
will be ignored. Further we will not even
tell the student. Our philosophy is that \emph{you can only know what
you know you know}, meaning that accidentally demonstrating
procedural abstraction not knowing that that is what you are doing
counts as nothing.

Each achievement will be marked as either \emph{pass},
\emph{fail} or \emph{fail with push-back}. A student that fails to
demonstrate an achievement is free to try again at her leisure.
The limited attempts plus the waiting time discourages students
from brute forcing their way to a pass by making continuous
attempts, accumulating feedback from the previous examiners to
present to the next. A fail with push-back indicates that the
student has misunderstood something deeply, prompting us to 
block any re-attempt for the rest of the lab session to encourage
spending some more time understanding that achievement.
There are no special limits on the number of attempts to demonstrate
one particular achievement.
Failures with or without push-back are not counted towards any grade.

An examiner's decision is
final, and the examiner does not have to motivate failing a
presentation. However, we ask examiners to explain and motivate
both passing and failing grades on demonstrations to reinforce
learning, build confidence, and to help students transition from a
failed demonstration to a passing one.

\subsubsection{Demonstrations \& Plagiarism}

Because the focus of a demonstration is the skill in relation to
the evidence, we feel that plagiarism is not really a problem. We
encourage students to discuss solutions, look at code written by
fellow students, and collaborate. Because
demonstration commonly involves examiners pointing at code, and
asking questions about it, or altering the code as part of the
discussion, students that lack a deep understanding of the code
have a hard time passing. Ultimately, the student is demonstrating
some abilities, not whether he or she wrote a particular piece of
code or not.

\subsection{Lectures}

The course uses traditional lectures. Many lectures have a
corresponding screen cast which covers similar content in a
different way, often with a focus on code, most of the time
showing an editor and a terminal and not slides. Several lectures
are given through ``live coding'' where the lecturer switches
between drawing on a whiteboard and projecting an editor and
terminal and incrementally constructing a program from scratch.
Each lecture is mapped to a set of achievements, to help students
(predominantly those without much prior programming experience)
make plans for what to demonstrate.
Following constructive alignment, mapping lectures to
achievements has been very helpful for lecture design---what
should be covered in this lecture, what can be cut in the event of
running out of time, etc.

\subsubsection{Discussion}
\label{sec:discussion}

Initially, students fear demonstrating mastery because they are not
used to the process, and find it difficult to understand
beforehand. To this end, we encourage students to attempt
presentations early on, not with the goal of passing, but with the
goal of understanding and practising how to give a demonstration.
This concept is eye-opening to many, and is easily explained in
terms of sacrificing an abundant resource (lots of demonstration slots remaining) in exchange for knowledge
that will be useful throughout the entire course. Usually,
once the initial resistance is broken down, students have no
problem requesting demonstrations.

One goal of reducing expected learning outcomes to much smaller achievements was
to reduce the expertise and amount of teacher training required to
be an examiner for demonstrations (especially judging quality more than pass/fail). The teaching staff on this course
is predominantly made up of junior TAs (bachelor or master
students) and senior TAs (PhD students). Because each achievement
demonstration is pass/fail (and push-back), we feel that it is
easy to deliver reasonably consistent quality both on judgement
and feedback. In combination with each student going through at
least 40--70 demonstrations, unfairness due to examiner
inconsistencies is not a problem in practice.

The focus on oral presentations put orally-skilled students at an
advantage, but this is mitigated by the evidence requirement and the
probing when examiners change code or introduce bugs. We
experience considerable improvement in presentation quality from
many students who initially are afraid, block, or otherwise are
uneasy when forced to talk in-front of others. We try to motivate
the use of oral examination by relating it to technical job interviews, and,
according to the course evaluation, by the end of the course most
students are happy with doing oral presentations.

\subsection{Working with Phases and Sprints}
\label{sec:working-with-phases}

We divide the course into three phases, one devoted to imperative
programming (currently in C), one to object-oriented programming
(currently in Java), and one devoted to a programming project. The
first two phases are further subdivided into two sprints each,
subdivision of the third phase is controlled by the students
themselves. Each sprint is two weeks and has one assignment. At
the end of each sprint is the \emph{soft deadline} for the
assignment. The soft deadline marks the time when an assignment
should be finished for the student to be perfectly on-time. At the
end of the following sprint is the \emph{hard deadline}, which is
the final deadline for handing in the assignment. Failure to meet
a hard deadline means that the student is derailing and is
subsequently called to a face-to-face meeting with the goal of
adjusting the plan for finishing the course. Again, there is no
punishment or points deducted, but students work hard to meet soft
deadlines and very hard not to break hard deadlines.
In fact, the idea of using deadlines as a tool to help with
planning was suggested by students after the first instalment of
the course (which had no deadlines).

As the course progresses, the number of assignments to choose from
in each sprint increases, reflecting our expectancy of student's
increasing maturity and ability to branch out.

\begin{figure}[ht]
  \centering
  \includegraphics[width=\textwidth]{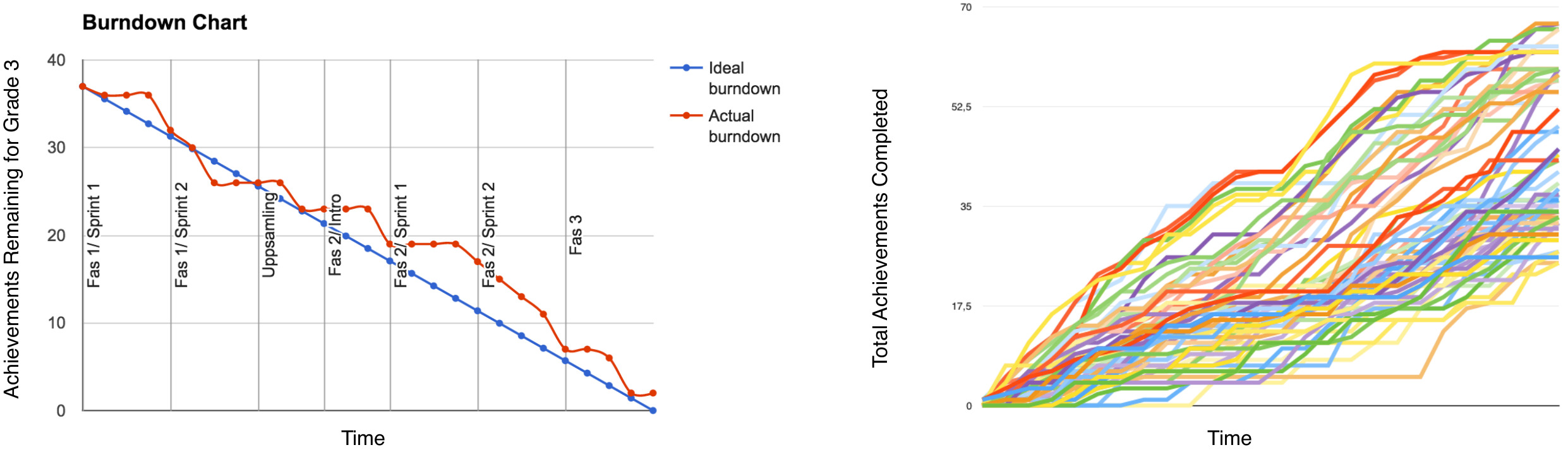}
  \caption{Left: a burndown charts for a student in 2016.
    The blue, straight line is the ideal burndown. The red line is
    actual burndown. The vertical labels are sprints and phases.
    Right: all students' achievements visualised as a burnup chart.
    It is clear that students' progress differ. (Colours have no meaning.)}
  \label{fig:burn}
\end{figure}

To reduce the administrative burden, we keep track of assignments
as achievements, meaning ``Successful demonstration of the Assignment
of Phase 1, Sprint 1'' is an achivement. For each sprint, students
are asked to plan for a number of achievements to demonstrate, in a
way similar to story points (see \eg{} \citet{cohn2004user}). We plot progress as remaining number
of achievements to desired grade and visualise this as a burndown
chart. We have experimented with several ways of visualising
student progress, most of which led to \emph{increased} stress.
Burndown charts tell a rich story, allow plotting trends,
exploring possibility of getting several grades at the same time, and
show periods of high/low activity etc. Figure \ref{fig:burn} shows an
example of burndown charts generated by students using a
provided on-line spreadsheet template. At the end of each sprint,
students meet a TA in groups of $\approx$10 to discuss their progress,
compare burndown charts and trade insights about assignments,
achievements, demonstrations, etc. These meetings serve as a
forcing function for updating the burndown chart and also force
students to openly discuss their (lack of) progress and make
plans. While some students are stressed by always chasing
the ideal burndown, they are also very happy about always knowing
what are the possible attainable grades and adjust their desired
grade up and down depending on a wide variety of reasons.

The division of the course into discrete units of time is
important to make progression through the course visible,
especially given that students are not required to make
demonstrations at any specific time. Judging from experiments that
we have made with more/fewer/longer/shorter sprints, this is a
quite flexible part of the design and finding a good
balance between a sense of urgency and student stress is key.


\subsection{The Coding Exam}
\label{sec:coding-exam}

The only guaranteed individual marking on the course is called the
coding exam, in which students use lab computers to solve
relatively simple programming problems without access to the Internet
in limited time. Cutting students off from the Internet breaks
the work-flow of many students who are used to getting unstuck by
searching and visiting web sites like Stack Overflow, or
discussing with a partner. We see that students must be
encouraged explicitly and repeatedly to practise solving problems
alone and---importantly---off-line, regardless of their programming skills.

While passing the coding exam is mandatory, it is strictly a
pass/fail examination and does not affect the final grade---its
only purpose is to present an obstacle that students will never be
able to get past without actually learning how to program (to some
limited degree). In our experience, many students have such low
confidence in their programming ability that they will avoid
touching the keyboard unless forced to. These students pair up
with other more confident students and let them sit in the driver position
throughout the course, and will learn many concepts shallowly as
is inevitable without any actual coding. We are completely open
with the students that the coding exam is an artificial obstacle
that is there to scare the students that would otherwise never
leave the passenger seat to take place in front of the keyboard.

Because the coding exam is done on lab computers, students have
access to the editors and IDEs that they use in the rest of the
course, and important tools like gdb and valgrind. We also
provide test cases that can be run easily from the command prompt,
allowing students to test their solutions. Ability to compile and
run test is a double-edged sword, however, as some students go
into panic mode when faced with hard error messages, and get stuck
on preliminaries such as forgetting what are the arguments to
\texttt{main} in C rather than focus on the parts that the exam
wanted to test.

To some extent, the code exam can be corrected automatically by
just running a set of tests on the hand-ins. Students passing the
code exam gets no other feedback than a passing grade. Hand-ins
for which the tests fail will be manually inspected and the code
handed back to the student annotated with a given a list of errors
and bad smells in the code, which corresponds to numbers in a
lengthy code exam report posted on the course web page.\footnote{Example (in Swedish, with lots of code examples): \url{http://wrigstad.com/ioopm/20161221.php}.} Because of automatic testing, results sans feedback can
usually be given within 24 hours of the code exam.
Only functional correctness determines pass/fail. This is partly to reduce
stress and help students to focus, but also because 
code quality comes up in almost all other parts in the course.

\subsection{Final Grade}
\label{sec:final-grade}

A student's final grade is decided solely by the number of
achievements successfully demonstrated. The achievements are
stratified after grade levels to match qualitative assessment that
could perhaps have been made using more comprehensive grading of
handed-in assignments, and in that sense ``quantifying quality.''
We do not allow, for example, substituting one achievement on level 5 for
another on level 4, etc. The achievements match the learning
outcomes decided by (in our case) higher powers, and when read in
that light, it does not make sense to substitute, for example, procedural
abstraction for knowledge of garbage collection.\footnote{A system
  identifying achievements that make sense to allow substitution
  can of course be constructed, but would add more complexity for
  questionable gain.} Students initially baulk at the concept of
getting 100\% (of anything) to get a corresponding grade, possibly
because of how other courses allow ``points trading.''

\section{(Self-)Evaluation}

In this section, we report on feedback from students (based on the
2016 non-mandatory course evaluation which asked 53 questions and
was filled out by 80 students), teaching assistants as well as our
own reflections. We concentrate on feedback on the achievement
system and not on the course on a whole. Feedback from teaching
assistants has not been collected through any systematic means.


\subsection{Feedback from Students}

In the 2016 course instalment, we felt that we had finally
understood how to explain the achievement system to students: by
introducing it as a set of checkboxes that they had to tick off at
some point that made
sense to them during the course. Students commented very positively on the
achievement system in the course evaluation, and suggestions for
change were generally in the form of (perceived) improvements such
as ``merge these achievements'', ``provide more suggestions for
clustering of achievements'' or ``clarify achievement X to make it
more clear''. Almost 3/4 of the students report that they grouped
achievements together for demonstration based on how the
achievements fit together. Almost 2/3 of students think that the
setup with constant small checks helped them avoid procrastinating,
though a little less than 1/3 disagree. Notably, 45\% of
students think that their stress-level increased due to the large
number of checks.

90\% of the students felt that it was always clear to them what
grades were attainable (66\% strongly, 24\% less strong). 75\% of
the students felt that the achievement system has helped them
absorb the contents of the course. Students generally perceive being more responsible for
driving their own studies (\emph{e.g.}, clustering achievements,
planning etc.) as something good.

1/3 of the students think that the oral examination has
\emph{greatly} influenced their ability to explain, motivate and
communicate. Another 51\% agree but to a lesser extent. There is a
strong preference for immediate oral feedback over delayed
feedback in written form.

40\% of the students do not think that they would have gotten more
out of the course by doing \emph{fewer} oral checks, though 26\% think
that they would have gotten a little more out of it. 18\% think
they would have gotten a lot more out of it. Notably, oral checks
get the most complaints due to the time students spend queuing,
waiting to demonstrate. Invariably in the last years, illness or
other unfortunate circumstances have caused a shortage of
examiners a few times each semester, causing waiting times to
inflate and students to complain. In 2016, the average waiting
time was 53 minutes between asking to demonstrate and having the
grade entered into the system (so this includes demonstration
time). To our mind, this is relatively short, but there are also
many demonstrations. Some students say that ``waiting time is not
a problem, since there is always something to work on while
waiting.'' Some students spend the time waiting in the queue
\emph{waiting}, which seems wasteful. In future work, we wish to
understand this behaviour --- is it because a looming
demonstration is stressful, because a pending demonstration is an
alibi to take a break, because of fear of swapping out a rehearsed
demonstration, etc.

\subsection{Feedback from Teaching Assistants}

An effect of allowing students to demonstrate any achievement at
any time is that teaching assistants need to be prepared to assess
a mastery demonstration about any part of the course material at
any time\footnote{In practice though, there are some natural
  constraints on which achievements are actually demonstrated. For
  example, achievements concerning object-oriented programming are
  typically not demonstrated until later in the course, and
  students often tend to demonstrate achievements that their peers
  have successfully demonstrated. }.
This is very different from other courses where students typically
all work with the same assignment at the same time. Senior TAs and
more experienced junior TAs see this as something positive with
the course, as the work becomes more stimulating and less
monotonic. Junior TAs are sometimes instead stressed by this,
especially early on in the course, because they feel like they
have to be able to answer any question about the course. To
mitigate this we encourage TAs to ask each other for help, and to
sit in and listen to each other's mastery assessments in order to
learn from each other. Throughout the course we also have
semi-regular lunch meetings together with the TAs to discuss what
works well and not, as well as trading ``assessment strategies''
for different achievements.
TAs that have taken the course themselves are typically more
comfortable with the assessment situation as they have experienced
it from the other side.

Many TAs are stressed when there is high pressure on the
demonstration request list and long waiting times in the lab
rooms. In these situations, TAs sometimes have a hard time
balancing the quality and quantity of assessments; giving more
detailed feedback means that other students have to wait even
longer for their chance to demonstrate.

\subsection{Our Own Reflections}

The main negative aspect of switching to the achievement-based
system is that a lot of time that was previously spent casually
chatting with students during lab sessions is now spent taking
demonstrations. This leaves less time for interested students to
discuss with teaching assistants, and less time to form stronger
connections with students. It is also likely that students spend
less time programming, and more time reflecting, rehearsing, and
demonstrating.

In our opinion, students learn \emph{more} and are more aware of
what they learn. Students are able to discuss concepts like
abstraction, modularity, inheritance, etc. at a much higher level
and relate these concepts to their own code. To a greater
extent than before, students also form \emph{opinions} about
concepts and constructs, rather than just accepting them as
decreed. To an overwhelming extent, students now think
of themselves as the active agents in their learning, in contrast
to previous years when students were expecting to \emph{be
  taught}. To a larger extent than before, students are aware of
their own strengths and weaknesses, and ability to plan and
execute plans. The ability to reflect on their own learning was
also observed by Matthias Hauswirth, a professor from Lugano
visited the course for a day in 2016 and carried out interviews
with several students.\footnote{Matthias Hauswirth, personal
  communication.}


\section{Summary \& Conclusion}

Four years ago, we radically changed our course on imperative and
object-oriented programming to a design where students are forced
to take responsibility for their own learning. By reifying the
formal course goals as a large number of small achievements to
``unlock'', we give the students a quantifiable measurement of
progress throughout the course. By separating the achievements
from the mandatory assignments, the students themselves get to
make the connection between the course concepts and their code.

In our experience, the students are able to explain and discuss
course topics at a higher level than before, and after the third
iteration of the course, we are at a point where a majority of
students are at least as satisfied with the course as they were
before the change. The course requires a large staff of teaching
assistants to handle the oral examinations, but because the course
produces students who are well-versed in the course contents,
there is never a shortage of junior teaching assistants to employ.
Scaling the course to accommodate a growing number of students in
the future requires employing more teaching assistants to keep the
waiting time in the lab sessions sufficiently low.

Because the course has changed in many ways (\eg{} doubled in
size) it is hard to say whether the course is more or less
labour-intensive than before switching to achievements. It is
budgeted the same way as all other courses at the department, and
fits in that budget. A cost-neutral difference is that we seem to
go through \emph{more} TAs each working \emph{fewer} hours than
before. Our guess is that this is because TA work is now more
intense as it is more oriented towards examination than before.


\subsection{Lessons Learned}

We end with some of the lessons we have learned while teaching
this course.

Over the years, we have reduced the number of achievements
slightly, mostly to make the list more manageable for students.
Achievements that are hard to group with other achievements were
the first to go.
Related to this point, changes to the achievement system is a pain
when dealing with old students, who did not pass the course (to some
degree or at all) in a previous year. It is
hard to translate the progress from one year to the other when the
set of achievements are different.
This problem can be alleviated by supporting partial grading. In
our course we give 25\% of all credits for completing 50\% of the
mandatory assignments (excluding the project) and $\approx$50\% of the achievements needed
for a passing grade (again, excluding achievements tied to the project). This way, students who fall far behind for
some reason can aim to finish \eg{} the assignments for the
imperative phase and finish the object-oriented phase from the
beginning the following year.

Due to the large number of demonstrations, computer support is
crucial, both for handling the queue of demonstration requests and
to keep track of the progress of each student. The large volume
also means that errors will happen, meaning that the backend needs
to support manual input of results.
The fact that everything is recorded digitally is also a great
source for data-mining. We can extract which achievements are
currently the most popular, which achievements most students fail,
statistics on waiting times etc. We can also report on the
progress of the entire class to help students understand how their
progress relates to the average progress of the class.

\subsection{Final Remark}

Achievement Unlocked is an interesting combination: it has elements of
mastery learning \citep{bloom1968learning},
flipped class-room \citep{bergmann2012flip} (in the sense of students doing lots of work at home and then come to discuss with teachers and TAs in lab sessions),
gamification (with respect to resource management and unlocking of grades),
and its design was heavily influenced by our belief in constructive alignment \citep{biggs1996enhancing}.
As course designs go, it is complex, but purposely so---with the
intent of forcing students to actively work with their learning
process and reflect on study techniques, but reasonably scalable to the
extent affordable junior assistants are available. 

\bibliographystyle{ACM-Reference-Format}
\bibliography{main}

\appendix

\section{Achievements at a Glance}

The list below gives a high-level overview of what the
achievements in the course are and how they are grouped together.
Remark that the grouping serves no other purpose than to help
students abstract and thus reduce the perceived volume of skills
to learn and demonstrate mastery of.

\begin{description}\addtolength{\itemindent}{-2em}
\item[Abstraction] Achievements in this group concern procedural
  and object-oriented abstraction, the importance of naming, and
  information hiding.

\item[Code Review] Informal code reviews, responding to code
  reviews, refactoring.

\item[Communication] Essay writing, giving a structured
  presentation, working as a teaching assistant. The last
  achievement is on the highest grade-level and forces students to
  work as TAs during half a lab session where they work through
  the help list, and the students they are helping give them
  feedback. Part of the reason for having the helped students
  producing feedback is so that the helper must think actively
  about what the goals are of helping---not solving their problem,
  but helping them coming up with the solution themselves.

\item[Documentation] The single achievement in this group deals
  with writing interface documentation for other programmers to
  use.

\item[Encapsulation] Aliasing, name-based encapsulation, nested
  and inner classes, interplay of strong encapsulation and
  testing.

\item[Generics] Dealing with generics both in Java and C, and
  designing with/for parametric polymorphism.

\item[Imperative Programming] Recursive vs. iterative solutions,
  tail recursion elimination.

\item[Inheritance] Object-oriented inheritance, overriding,
  overloading, subtype polymorphism, separation of concerns.

\item[Memory Management] Allocation on stack vs. heap, manual
  memory management, automatic memory management, manual vs.
  automatic memory management.

\item[Methodology] Defensive programming, exception handling,
  failure management and fault tolerace.

\item[Modularisation] Module boundaries and interfaces, coupling
  and cohesion, separation of concerns.

\item[Object, identity \& structure] Identity vs. equality, value
  semantics, concrete vs. abstract classes.

\item[Planning] The single achievement in this group deals
  exclusively with planning and following-up, which is covered in
  Section \ref{sec:working-with-phases}.

\item[Pointers] Pointers and arrays in C, pointer-based linked
  structures, pointer semantics, indirection and pointers to
  pointers.

\item[Pragmatics] Compilers, interpreters, JITing, linking, and
  binding.

\item[Profiling \& Optimisation] Profiling for performance and
  memory usage, optimising performance and memory usage guided by
  profiling results.

\item[Testing] Unit testing, test quality, static anlysis,
  fuzzing.

\item[Tools] Debuggers, documentation tools, editors, build tools,
  working the terminal.

\item[Assignments] This group collects the achievements mapping
  directly to assignments (\eg{} Assignment in Phase 1, Sprint 1,
  etc.).

\item[Project] This group collects the achievements that are part
  of the programming project at the end of the course. These are
  not demonstrated in the usual way, but we unify them with the
  other assignments for simplicity. Project achievements include
  using and evaluating the usefulness of pair programming,
  regression testing, working with pull requests and code reviews,
  coding standards, etc.

\end{description}

\subsection{Example Achievement: 
  Implementing Generic Data Structures with Void Pointers}

\noindent
\textbf{Grade level: 3} \\
\textbf{Demonstrated: in lab sessions} \\[1ex]
\emph{Use void pointers in C programs in relevant ways to
    implement genericity, for example, a collection capable of
    storing any kind of data.}\\[1ex]
  Multiple copies of the same code with minor differences is
  undesirable. Consider for example two separate implementations
  of lists of integers and lists of floats where the second was
  created from the first. If a bug is found in one list, two sets
  of bug fixes must be carried out in equivalent ways to make sure
  the lists' behaviour stays identical.\\[1ex]
  C supports pointers to data of unknown type (void pointers,
  \texttt{void *}) which can be used to build a general list (for
  example). Constructing a list where elements are void pointers
  allows the list to hold arbitrary data.\\[1ex]
  Void pointers is a common C idiom that is also used in many
  places in the standard C library to identify a memory location
  with unknown content. The compiler does not know what is at
  the location, nor the size of the object at the location.\\[1ex]
  Demonstrate your understanding of the void pointer idiom, and
  how the compiler's lack of knowledge of the type of the object
  at an address influences programming. Start your demonstration
  by explaining the concept of genericity, how and when it is
  useful, and what you think about C's support for genericity.

\section{Course Time-Line}
\label{outline}

With few exceptions, each week has two 2-hour lectures and three
scheduled 4-hour sessions for demonstration (and getting help).

\begin{description}
\item[C Bootstrap (2 weeks)] (Broad but not deep): loops, recursion,
  basic I/O, functions, arrays, pointers, strings, modularisation
  and separate compilation, typedefs and basic data types,
  structs, unions, function pointers. What is good code? \\[.5ex]
  All C assignments are made available at the end of the bootstrap.  
  All Java assignments are made available in C Sprint 1 or early in C Sprint 2.
  Each sprint has one assignment and the soft deadline for the assignment
  is at the end of that sprint. Hard deadline at the end of following sprint. \\[.5ex]
  There are five mandatory labs in the bootstrap. Soft deadline on
  the same day as the lab is scheduled. Hard deadline on the
  following lab.
  
\item[C Sprint 1 (3 weeks)] (Students build a small interactive program.)
  Recaping from the bootstrap. Imperative vs. functional
  programming. Stack, heap and manual memory management. Pointers.
  Top-down and bottom-up design of programs. Layer design.
  Scripting and automation. Basic testing. What is a good
  programmer?

\item[C Sprint 2 (3 weeks)] (Students rebuild parts of the program in more
  dynamic ways, eg. replacing arrays with trees, carve out
  reusable libraries and swap with each other.) Linked structures
  and iterators. Macros and the C preprocessor. Defensive
  programming. Modules, coupling and cohesion. Readable code.
  Bit manipulation. Profiling and optimisation. 

\item[Java Bootstrap (1 week)] Objects and classes. Basic Java.
  What parts of C carry over to Java. Automatic memory management.
  Execution environments (VMs, JITs, etc.)

\item[Java Sprint 1 (3 weeks)] OOAD. Constructors. Initialisation. References. 
  Inheritance. Overloading and overriding. Abstract classes. 
  Interfaces. \\[.5ex]
  Students choose between a classic OO simulation (cashiers and
  waiting lines in a store), or a very simple micro blogging
  application with a text-only interface that uses the network.
  (Lots of code is provided for the latter.)
  
\item[Java Sprint 2 (3 weeks)] Parametric polymorphism. Testing. Static and
  dynamic binding. Profiling. Refactoring. Java Bytecode. 
  Garbage collection.\\[.5ex]
  Students choose between implementing a simple text-oriented MUD
  or a symbolic calculator. Both assignments focus on object structures,
  interfaces, inheritance and polymorphism. 

\item[Project (4 weeks)] Basic Scrum.\\[.5ex]
  The project implements a basic automatic memory management system
  for simple C programs (a conservative Bartlett-style collector).

\end{description}

\end{document}